\documentclass[preprint,prl]{revtex4}

\input{tcilatex}
\begin{document}

\title{Field cooling memory effect in Bi2212 and Bi2223 single crystals.}
\author{D. Shaltiel$^{1}$}
\author{H-A. Krug von Nidda$^{2}$}
\author{B. Rosenstein$^{3}$}
\author{$^{+}$B. Ya. Shapiro$^{4}$}
\author{M. Golosovsky.$^{1}$}
\author{I. Shapiro$^{4}$}
\author{A. Loidl$^{2}$}
\author{B. Bogoslavsky$^{1}$}
\author{T. Fujii$^{5}$}
\author{T. Watanabe$^{6}$}
\author{T. Tamegai$^{7}$}
\affiliation{$^{1}$Racah Institute of Physics, The Hebrew University of Jerusalem,
Jerusalem, Israel. $^{2}$Experimental Physics V, Center for Electronic
Correlation and Magnetism, University of Augsburg,86135 Augsburg, Germany. $%
^{3}$National Chiao Tung University, Department of Electrophysics Hsinchu,
Taiwan. $^{4}$Department of Physics, Bar-Ilan University, Ramat-Gan, 52100,
Israel. $^{5}$Cryogenic Research Center, The University of Tokyo, Yayoi,
Bunkyo-ku, Tokyo 113-0032, Japan. $^{6}$School of Science and Technology,
Hirosaki University, Hirosaki, Aomori 033-8561, Japan. $^{7}$Department, of
Applied Physics, The University of Tokyo, Hongo, Bunkyo-ku, 113-8656 Japan.}

\begin{abstract}
A memory effect in the Josephson vortex system created by magnetic field in
the highly anisotropic superconductors $Bi2212$ and $Bi2223$ is demonstrated
using microwave power absorption. This surprising effect appears despite a
very low viscosity of Josephson vortices compared to Abrikosov vortices. The
superconductor is field cooled in DC magnetic field $H_{m}$ oriented
parallel to the $CuO$ planes through the critical temperature $T_{c}$ down
to 4K, with subsequent reduction of the field to zero and again above $H_{m}$%
. Large microwave power absorption signal is observed at a magnetic field
just above the cooling field clearly indicating a memory effect. The
dependence of the signal on deviation of magnetic field from $H_{m}$ is the
same for a wide range of $H_{m}$ from $0.15T$ to $1.7T.$
\end{abstract}

\maketitle

Many complex disordered systems in nature fail to equilibrate with their
environment below certain temperature. Examples include various glasses,
from window glass to spin glass \cite{glass1}. The glassy metastable
frustrated state gives rise to novel and unusual behavior such as memory
effects, ageing and nonlinear dynamics. In recent years the glass family has
significantly expanded and now it includes additional effects such as domain
walls \cite{Repain}, two dimensional electron gas \cite{Vaknin} and
Abrikosov vortices in type II superconductors under magnetic fields (see 
\cite{Andrei} and references therein).

The glassy vortex state is usually prepared by field-cooling (FC) the sample
containing quenched disorder across the superconducting transition
temperature $T_{c}\left( H\right) $. A memory phenomenon in superconductors
is usually studied as a response to an applied perturbation which suddenly
disturbs the superconductor. After removing the perturbation, subsequent
response shows the memory effects, i.e. the rate of the magnetic relaxation,
for example, is slower in "older" systems than in "younger" ones \cite%
{Andrei}. Experiments on different conventional superconductors \cite%
{Henderson} and in Y-Ba-Cu-O single crystals \cite{Valenzuela} demonstrated
the presence of a history effect similar to those in Heisenberg spin glasses.

Here we report an entirely different manifestation of the memory effect in
strongly anisotropic high $T_{c}$\ superconductors. In these
superconductors, magnetic field directed parallel to the layers penetrates
the sample as a system of Josephson vortices (JV) qualitatively distinct
from Abrikosov vortices (AV) appearing when the magnetic field is
perpendicular to the layers (or in less anisotropic materials like in most
conventional superconductors and optimally doped $YBaCuO$). It is well known
that for sufficiently strong disorder the AV\ form vortex glass \cite%
{Blatter}. In JV system a randomly distributed set of defects of the layered
structure (microresistences \cite{Barone}), is analogous to the quenched
disorder pinning AV in the conventional vortex glass. Apparently the pinning
of JV is expected to be less effective than the pinning of the AV since the
JV are coreless. However as it was shown (see \cite{Barone}) these JV can be
pinned. \qquad \qquad

In this note we demonstrate the formation of the glassy state of the JV \cite%
{Barone} in $Bi_{2}Sr_{2}Ca_{1}Cu_{2}O_{\delta }$\ ($Bi2212$) and $%
Bi_{2}Sr_{2}Ca_{2}Cu_{3}O_{\delta }$($Bi2223$) single crystals. The memory
properties are studied by measuring the microwave power absorption (MPA) of
the sample subjected to colinear DC and AC magnetic field\textit{s} parallel
to the layers. This method is sensitive to register weakly pinned JV, while
strongly pinned JV do not contribute to the signal \cite{Shaltiel}. In the
present experiment the JV glass is prepared by field cooling (FC) the sample
from above $T_{c}$\ to $T_{m}=4.2K$\ subjected to the DC magnetic field $%
H_{m}$\ parallel to the $a-b$ crystal plane. At this temperature\ the
external field is decreased to zero and after applying a low frequency AC
field \ it\emph{\ }is subsequently increased again . No MPA signal is
detected until the external magnetic field reaches the initial field $H_{m}.$
The peak of the MPA signal appears just above $H_{m},$ demonstrating the
memory effect. The dependence of the signal on deviation of magnetic field $%
H-H_{m}$ is the same \textit{in} the wide range of the cooling fields.

The memory properties of the Josephson glass are investigated using a Bruker
ELEXSYS continuous wave electron spin resonance (ESR) spectrometer\ working
at X-band frequency, $\omega _{mw}/2\pi =9.3GHz$, see ref. \cite{Shaltiel1,
Shaltiel} for details. A microwave source feeds a rectangular H102 cavity
where a optimally grown $Bi2212$ or $Bi2223$ crystal (sizes $1\times 1\times
0.1mm^{3}$)\ is placed in the center of the cavity region where only
microwave magnetic field is present. The microwave magnetic field was
parallel to the $a-b$ plane. The sample whose temperature was varied
continuously down to helium temperatures using a helium continuous flow
cryostat (Oxford Instruments) was exposed to colinear DC and AC ($\omega
_{AC}/2\pi )=100kHz)$ magnetic field parallel to the a-b plane and
perpendicular to the microwave field. The microwave reflected from the
cavity is rectified by a diode that feeds a lock-in detector. The presence
of a small AC component significantly increases the magnitude of \ the MPA
signal, due to vortex shaking \cite{Shaltiel}. \ The technique allows
measurement of the electronically detected peak to peak difference of the
MPA signal 
\begin{equation}
S\left( H_{DC}\right) =max_{t}P-min_{t}P\ \approx dP/dH_{DC}  \label{na1}
\end{equation}%
\ \ as a function of the $DC$\ magnetic field, $H_{DC},$\ parallel to the
layers.

The high quality $Bi2212$ crystal used in the present study was grown by the
floating zone technique using an image furnace \cite{Tamegai}. Crystal from
the same source have been used in various investigations of the vortex
matter before \cite{Zeldov, Bending}. The Abrikosov ($H||c$-axis) vortex
physics in the $Bi2212$, whose anisotropy is $\gamma =250$ \cite{Shimoya},
was comprehensively studied \cite{Zeldov}. A complicated $H$-$T$ phase
diagram contains the melting line and a transition to a glassy state. For a
much more complicated case of tilted fields in which both the Abrikosov
("pancake") vortices and JV are present the melting line was established
recently \cite{Konczykowski}. However the glassy phase for JV, even in the
simplest case of $H\perp c$-axis has not been studied so far and the
microwave technique enables to approach it.

In the less anisotropic material $Bi2223,$ $T_{c}=110K$ and $\gamma =60$ 
\cite{Shimoya}, the JV physics in this compound has not been extensively
studied. Although the $Bi2223$ crystal had inherent defects due to
intercalations of other phases (that are always present \cite{Liang}), the
memory effect is only slightly affected compared to that in $Bi2212$ .

Microwave absorption\ in the presence of JV in the layered superconductors
has been studied theoretically \cite{Koshelev}. In the vortex state created
by the DC field and carrying a microwave supercurrent $J_{s}=J_{s0}\exp
\left( -i\omega _{mw}t\right) $ along the $c$-axis, the MPA in weak magnetic
field is described by the single JV\ formula \cite{Shaltiel}:\textbf{\ } 
\begin{equation}
P=\frac{2\nu _{ab}}{\omega }\frac{h}{k+h^{3}}.  \label{na5}
\end{equation}%
Here $h$ is the DC magnetic field in units of the Josephson critical field $%
\varphi _{0}/2\pi \lambda _{c}\lambda _{ab},$ $\omega =\omega _{mw}/\omega
_{p}$ with $\omega _{p}$ being the plasma frequency$\ $and $\nu _{ab}=4\pi
\sigma _{ab}\lambda _{ab}^{2}\omega _{p}/c^{2},$where $\sigma _{ab}\ $and$\
\lambda _{ab}\ $are the\ normal$\ state$ conductivity\ and\ penetration\
length\ in\ the $ab$\ plane\ correspondingly. The constant $k$\ is 
\begin{equation}
k=\frac{\left( K/\rho _{J}-C_{c}\right) ^{2}}{2\pi \omega ^{4}}+\frac{\nu
_{ab}^{2}C_{ab}}{2\pi \omega ^{2}},  \label{n6}
\end{equation}%
where coefficients $\rho _{J}$ and $\eta _{J}$ are the mass and the
viscosity of the JV calculated by Koshelev \cite{Koshelev} and $%
C_{c}=9,C_{ab}=2.4.$

To study irreversible properties (memory effects) we execute a field cooling
protocol presented in Fig.1. The samples are field cooled in a DC field $%
H_{DC}=H_{m}$\ parallel to the $a-b$ plane from above $T_{c}$\ (point $A$ in
Fig.1) to a low temperature $T_{m}$ (point $B$).\ Unless otherwise
specified, $T_{m}$\ is equal to $4K$. Then the field is rapidly reduced to
zero (point $C$), after which it is raised again to above $H_{m}$ (point $D$%
). A characteristic peak is observed slightly above $H_{m}$, indicating that
it results from a memory effect.

The signal intensity in $Bi2212$ as function of the DC magnetic field in the
range $0.15T$\ to $1.7T$\ ($1.7T$ is slightly below the limiting field of
the spectrometer) is presented in Fig. 2a. The height and position relative
to cooling field $H_{m}$ are the same for magnetic fields in the wide range,
as demonstrated in Fig. 3. Their sharp maximum on this plot appears at about 
$0.05T$\ indicating that the memory signal occurs at the same field at a
value slightly above $H_{m}$\ for all cooling fields.\ The dependence of the
signal intensity and the signal shape for different cooling temperatures are
presented in Fig. 4 for $H_{m}=0.3T$. The effect persists even if the
cooling temperature was increased to $10K$, but disappears at about $15K.$%
This indicates that the glass transition temperature $T_{g}\left( H\right) ,$
line 2 in Fig.1, is between $10K$ and $15K$ depending weakly on magnetic
field. We found that in the range of fields studied, that the precursor of
the glass transition appears at much higher temperatures around $30K.$In
this crossover region the memory effect is not fully developed, see $T=10.5K$
line in Fig. 4.

Experimental results similar to those presented in Figs. 2a for $Bi2212$
were observed in a less anisotropic $Bi2223$ superconductor which was field
cooled to 4K. In Fig 2b MPA intensity as function of the DC field are
presented for magnetic fields $H_{m}$\ up to $0.25T$.

The glassy behavior observed in JV systems in BSCCO for the first time is
reminiscent of those observed in other glassy systems. Specifically it is
likely that the origin of JV pinning in the BSCCO is the "microresistance"
appearing \ due to oxygen deficiency along the layers \cite{Barone}. It
seems th We therefore apply theoretical ideas used to describe the glassy
state. One can explain the memory effect described above using the
hierarchical model of glassy state commonly used in the spin glass theory 
\cite{glass2}. When the sample is adiabatically field cooled (cooling speed
around $10$ $K/s$) from above $T_{c}$ to a low temperature $T_{m}$ well
below $T_{c}$ the vortices explore the available phase space and nucleate
into a deepest valleys of the landslide potential forming a quasi
equilibrium JV glassy state. The cooling in the present experiment is
definitely slow enough and temperature high enough, so that the most
favorable pinning positions are effectively found. When the external DC
magnetic field is dropped to zero, while holding temperature at $T_{m}$, the
vortices are still confined inside the deep pinning landslide minima and
cannot effectively absorb the microwave radiation. As the field is
subsequently increased (with the temperature still kept at $T_{m}$), the
fluxons remain immobile and prevent penetration of new JV till its starting
"equilibrium" field $H_{m}$ is reached.\ When the field is further increased
beyond $H_{m}$\ the JV glassy state looses its quasi equilibrium and new
vortices penetrate the sample populating unoccupied shallower valleys of the
pinning landscape (in which the pinning constant is smaller). These are the
vortices responsible for experimentally observed sharp MPA signal. It is
important to emphasize that the MPA signal depends on the magnitude of the
external magnetic field rather than on magnetic induction which is dominated
by the magnetic field of the pinned vortices (hystheresis). Thus the present
memory signal has the universal shape (see \cite{Shaltiel}, \cite{Koshelev}%
). The shape can be understood from Eqs.(\ref{na5},\ref{n6}) under an
assumption of the relatively small AC field amplitude (of order $0.001T$),
which allows expansion. In this case the peak to peak signal has a simple
form,%
\begin{equation}
S=\frac{2\nu _{ab}}{\omega }\frac{\left\vert k-2\left( h_{DC}-h_{m}\right)
^{3}\right\vert }{\left[ k+\left( h_{DC}-h_{m}\right) ^{3}\right] ^{2}},
\label{n7}
\end{equation}%
that for $h_{DC}-h_{m}>\left( \kappa /2\right) ^{1/3}$, demonstrates a well
pronounced scaling behavior in the coordinate $\left\vert
H_{DC}-H_{m}\right\vert $ of the Fig. 3.

The memory should be destroyed at higher temperature $T>T_{g}$ since the
pinning potential will be smeared (thermal depinning). This is the "glass
transition" in the JV system. Since the basic mechanism is the single vortex
pinning (for not too high fields) it is clear that the glass transition
temperature $T_{g}$ is only weakly dependent on field. The situation is not
expected to change qualitatively at larger fields in which JV will be pinned
collectively.

To summarize: although the "memory effect" in the context of vortex matter
has been extensively discussed, our experiments allow its demonstration in a
qualitatively different Josephson vortex system. The highly anisotropic
compounds $Bi2212$ or $Bi2223$, when field cooled to 4K, an MPA signal is
retrieved at a magnetic field close to the cooling field clearly indicating
a memory effect. The most striking observation in the "memory effect" is
that the microwave absorption peak obeys a universal scaling and appears at
the field just above the sample was equilibrated upon field cooling in
magnetic fields ranging from close to zero up to at least $1.7T.$

\begin{acknowledgments}
\begin{acknowledgments}
\bigskip \textit{Acknowledgements}
\end{acknowledgments}

\begin{acknowledgments}
We thank Zakir Seidov for experimental assistance. This work was supported
by The Israel Science Foundation, Grant No 499/07, by Heinrich Hertz Minerva
Center for High Temperature Superconductivity, by the German
Bundesministeriu fuer Bildung und Forschung BMBF under Contract No VDI/EKM
139617, by the Deutsche Forschungsgemeinschaft DFG within
Sonderforschungsbereich SFB 484 Augsburg, and partly supported by
Grant-in-Heinrich Hertz Minerva Center for High Temperature
Superconductivity, by the German Bundesministeriu fuer Bildung und Forschung
BMBF under Contract No. VDI/EKM 13N6917, by the Deutsche
Forschungsgemeinschaft DFG within Sonderforschungsbereich SFB 484 Augsburg.
It is partly supported by Grant-in- Aid for Scientific Research from the
Ministry of Education, Culture, Sports, Science, and Technology, Japan. B.R.
is supported by DOE and NSC of R.O.C. Grant No. M98-2112- M009-023. One of
us D. S. would like to thank O. Shaltiel for analyzing the experimental data
and M. Bezalel for his help.
\end{acknowledgments}
\end{acknowledgments}

\bigskip \newpage Figure captions.

Fig. 1. The field cooling protocol $A\rightarrow B\rightarrow C\rightarrow D$
on the $H-T$ phase diagram. Memory effect does not exists above the line 1
(reversible or JV liquid state). A precursor appears below this line and
becomes a well pronounced below the glass line 2.

Fig. 2a. Signal intensity as a function of DC magnetic field in $Bi2212$
crystal for different $H_{m}$ values up to $1.7T$, obtained by field cooling
at $H_{m}$ from above $T_{c}$ to 4K and then dropping the field to zero. The
signal's maxima are slightly above $H_{m}$, implying a memory effect.

Fig. 2b. Signal intensity as a function of DC magnetic field in $Bi2232$
crystal for different $H_{m}$ values up to $0.25T$.

Fig. 3. The memory signals shown in Fig. 2a, as function of the difference $%
(H_{DC}-H_{m})$. It shows that the signals are superimposed, with their
maxima at about $0.05T$. The intensities and shapes are similar. The insert
is a theoretical fit for $H_{m}=0.15T$.

Fig. 4. Signals as a function of DC magnetic field cooled at $0.3mT$ to $4K$%
, $9K$, $10.5K$ and $15K$. The similar signal intensity and shape at $4K$
and $9K$ and the signal broadening at $10.5K$, indicates a glass phase
transition at about $10K$.

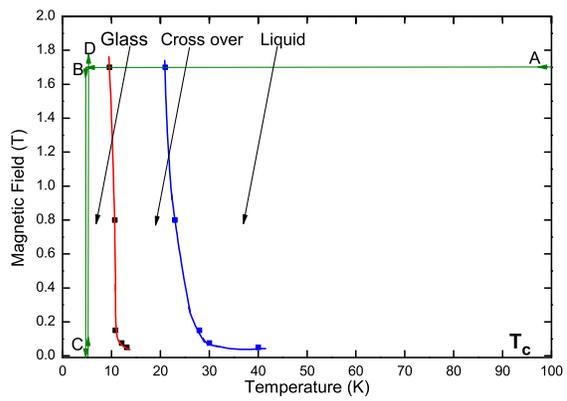
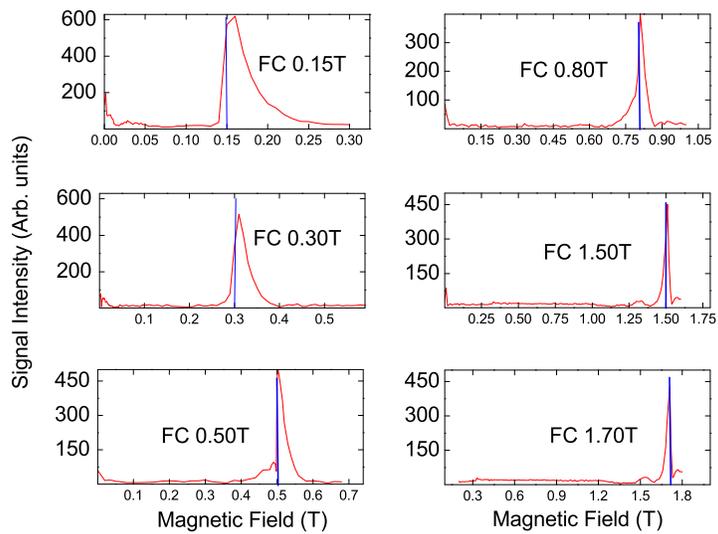
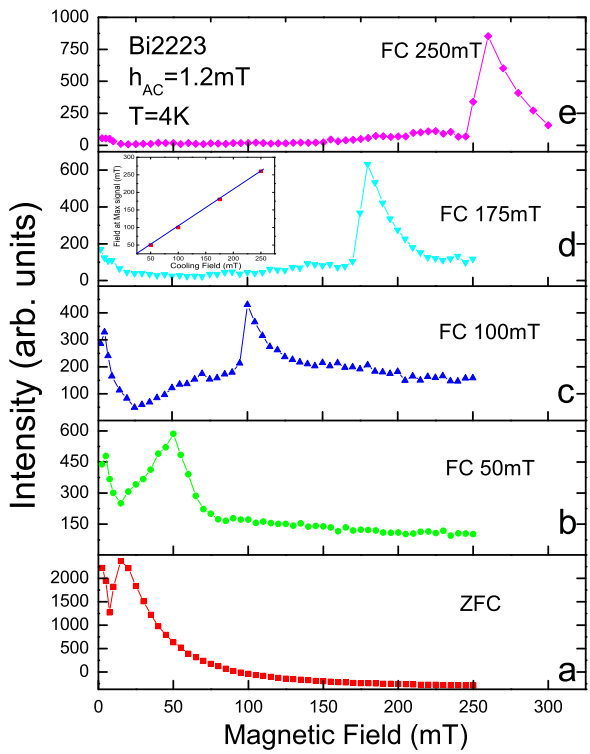
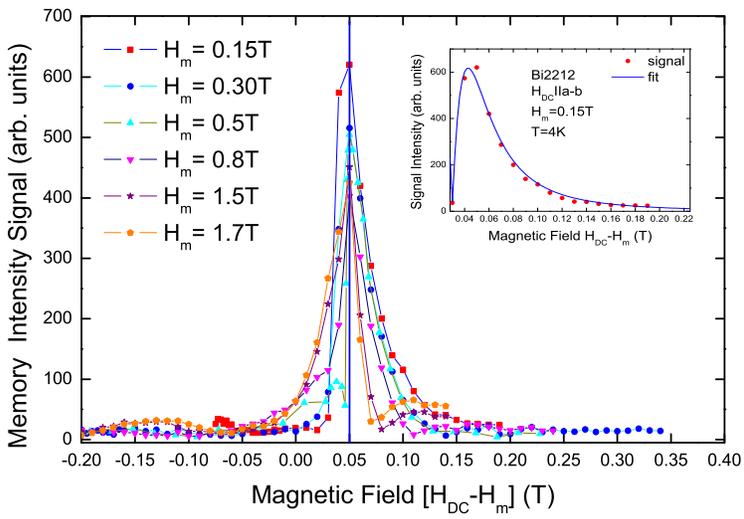
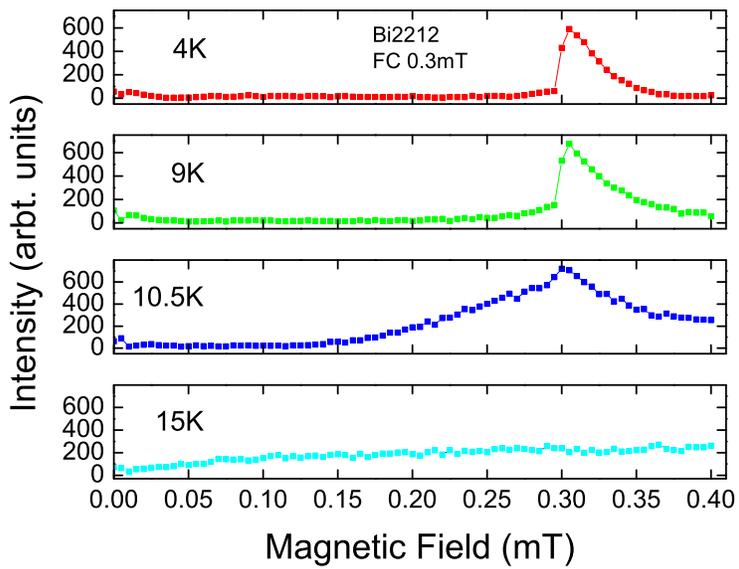
\end{document}